\documentclass[aps,showpacs,preprintnumbers,amsmath,amssymb,floatfix,nofootinbib,notitlepage]{revtex4-1}
\usepackage{graphicx}
\usepackage{ifpdf}
\usepackage{color}
\usepackage{colordvi}
\usepackage[normalem]{ulem}
\usepackage{slashed}
\usepackage{wasysym}
\usepackage{hyperref}
\usepackage[table]{xcolor}
\definecolor{nicered}{rgb}{0.7,0.1,0.1}
\definecolor{nicegreen}{rgb}{0.1,0.5,0.1}
\hypersetup{colorlinks,citecolor= nicegreen,linkcolor= nicered}

\usepackage{pifont}
\newcommand{\beqn}{\begin{eqnarray}}
\newcommand{\eeqn}{\end{eqnarray}}
\newcommand{\be}{\begin{equation}}
\newcommand{\ee}{\end{equation}}
\newcommand{\bal}{\begin{align}}
\newcommand{\eal}{\end{align}}

\newcommand{\ba}{\begin{array}{c}}
\newcommand{\bat}{\begin{array}{cc}}
\newcommand{\ea}{\end{array}}

\newcommand{\bi}{\begin{itemize}}
\newcommand{\ei}{\end{itemize}}

\usepackage{pifont}%

\definecolor{sexyred}{RGB}{220,20,60}

\newcommand{\TeV}{\, \mbox{\rm TeV}}

\def\beq{\begin{equation}}
\def\eeq{\end{equation}}

\begin{document}

\title{Null Hypothesis Test for Anomaly Detection}

\author{Jernej F. Kamenik}
\email{jernej.kamenik@cern.ch}
\affiliation{Jo\v zef  Stefan  Institute,  Jamova  39,  1000  Ljubljana,  Slovenia}
\affiliation{Faculty  of  Mathematics  and  Physics,  University  of  Ljubljana, Jadranska  19,  1000  Ljubljana,  Slovenia}
\author{Manuel Szewc}
\email{manuel.szewc@ijs.si}
\affiliation{Jo\v zef  Stefan  Institute,  Jamova  39,  1000  Ljubljana,  Slovenia}

\begin{abstract}
We extend the use of Classification Without Labels for anomaly detection with a hypothesis test designed to exclude the background-only hypothesis. By testing for statistical independence of the two discriminating dataset regions, we are able to exclude the background-only hypothesis without relying on fixed anomaly score cuts or extrapolations of background estimates between regions. The method relies on the assumption of conditional independence of anomaly score features and dataset regions, which can be ensured using existing decorrelation techniques. As a benchmark example, we consider the LHC Olympics dataset where we show that mutual information represents a suitable test for statistical independence and our method exhibits excellent and robust performance at different signal fractions even in presence of realistic feature correlations. 
\end{abstract}

\maketitle

%
\section{Introduction}\label{sec:intro}

The combination of increased experimental sensitivity and no clear leading theoretical guide for how physics beyond the standard model would manifest in current and future particle physics experiments has resulted in increased development of anomaly detection techniques for collider applications, see Ref.~\cite{hepmllivingreview} for a living review with a continuously updated list of references. These techniques, which make use of state of the art unsupervised and/or weakly supervised algorithms, have the advantage of being sensitive to a large variety of signals at the expense of losing statistical power in comparison to dedicated searches. However, appropriately quantifying said sensitivity is still an open problem~\cite{Shanahan:2022fzy}, with differing proposals, see e.g. Ref.~\cite{Park:2022zov}. An especially pressing question is how to evaluate the null hypothesis exclusion sensitivity of an anomaly detection method. The current strategy is to perform cuts using the anomalous score and extrapolate a background model from a control region. This can be problematic for several reasons. First, the use of the anomalous score itself to select events is not guaranteed to yield a robust method that disentangles the underlying processes, see e.g. Ref.~\cite{Kasieczka:2022naq} for a recent discussion of how ambiguities in the data representation can lead to different notions of anomalous events which vary in their discriminating power. Second, even if the anomaly score is an appropriate event selection tool, the use of cuts, which in an unsupervised search cannot be optimized on a targeted signal model, necessarily introduces a loss in sensitivity by discarding possible signal events. Finally, the use of a control region potentially introduces additional biases when assuming the absence of signal in the control region and/or employing interpolation methods such as the fit to a monotonic mass spectrum in a Bump Hunt.

In this work we aim to address some of the shortcomings outlined above. In particular, we propose a null hypothesis statistical test for anomaly detection which does not rely on fixed anomaly score cuts nor requires background model extrapolations from control regions.
We apply it to a specific anomaly detection technique, Classification Without Labels (\textsc{CWoLa}) introduced as a quark/gluon tagger in Ref.~\cite{Metodiev:2017vrx} and as an anomaly detection technique in Refs.~\cite{Collins:2018epr,Collins:2019jip}, and its extension introduced in Ref.~\cite{Benkendorfer:2020gek} incorporating simulation assisted decorrelation of features. We show that by testing for independence between the set of features used in the anomaly score, and those used to define signal and control regions, we can obtain a p-value which avoids false signal-detection and is robust in presence of slight correlations between the two sets of features. 

The work is structured as follows. In Section~\ref{sec:method} we review \textsc{CWoLa} and introduce the proposed statistical test. In Section~\ref{sec:application} we apply our method to a LHC Olympics benchmark to demonstrate its power and limitations. We conclude in Section~\ref{sec:discussion} where we also discuss possible future extensions and improvements. All the necessary code to reproduce our results is available at {\tt GitHub}~\cite{code}.

\section{Method}\label{sec:method}

Introduced in Ref.~\cite{Metodiev:2017vrx}, \textsc{CWoLa} is a weakly-supervised technique for anomaly detection which aims to learn a monotonic function of the Likelihood Ratio between Signal $S$ and Background $B$ processes for a set of features of interest $\vec{x}$, $\mathcal{L}_{S/B}(\vec{x})=p(\vec{x}|S)/p(\vec{x}|B)$, with the help of an additional feature $y$ uncorrelated with $\vec{x}$. The latter variable, often but not necessarily the invariant mass of the event, can be used to define two regions of interest: the signal region $M_{1}$ and the control (or side-band) region $M_{2}$, where the signal-to-background ratio is assumed to be higher in $M_{1}$ than in $M_{2}$. A weakly-supervised algorithm, \textsc{CWoLa} trains a classifier to distinguish between $M_{1}$ and $M_{2}$. The obtained output function $s(\vec{x})$ can then be mapped to $\mathcal{L}_{M_{1}/M_{2}}(\vec{x})$ through the likelihood ratio trick. The orthogonality of $y$ and $\vec{x}$ guarantees that $\mathcal{L}_{M_{1}/M_{2}}(\vec{x})$ is a monotonous function of $\mathcal{L}_{S/B}(\vec{x})$ and thus possesses in principle optimal statistical power.

Usual applications of \textsc{CWoLa} use the learned optimal classifier $s(\vec{x})$ to select events of interest and assign a certain significance to the difference in selected events in $M_{1}$ and $M_{2}$. The difference in the resulting selection efficiencies $\epsilon_{M_{1,2}}$ is a smoking-gun for the presence of signal in $M_{1}$ (and also $M_{2}$). However, this is only true in the limit of infinite statistics. In a realistic setting where the dataset is finite, quantifying the degree to which the difference in efficiencies relates to the presence of signal is non-trivial. One common strategy is to assume that there is no signal in $M_{2}$ and assess the agreement between the selected events in $M_{1}$ and a background extrapolation from $M_{2}$.

Our method constitutes an alternative to assess how the learned output $s(\vec{x})$ encodes differences between $M_{1}$ and $M_{2}$ caused by the presence of a signal. To introduce it, we focus on the density estimation framing of \textsc{CWoLa}, which clearly defines a background-only or null hypothesis. At its heart, \textsc{CWoLa} is a mixture model where $\vec{x}$ and $y$ are assumed to be conditionally independent given the process label $z=\{S,B\}$. After defining $M_{1}$ and $M_{2}$ using $y$, the trained classifier output is a function $s(\vec{x})$ {that inherits the conditional independence with respect to $y$}. The statistical model can be explicitly written as
\begin{equation}
    p(s(\vec{x}),y|\pi)=(1-\pi) \, p(s(\vec{x})|B)p(y|B)+\pi \, p(s(\vec{x})|S)p(y|S) \,,
\end{equation}
where $\pi$ is the signal probability. The background-only hypothesis is explicitly written as $p(s(\vec{x}),y|\pi=0)$ and corresponds to the case where the observed data shows independence between $s(\vec{x})$ and $y$. This is the key observation for our strategy. For a given measured dataset of pairs $\{s(\vec{x_{i}}),y_{i}\}$, one can assess whether they are statistically independent. If statistical independence is ruled out, the background-only hypothesis is ruled out, provided conditional independence holds. Conversely, if statistical independence cannot be ruled out, one has a clear statement about the incapability of \textsc{CWoLa} to discern whether any difference between $M_{1}$ and $M_{2}$ originates from the presence of a signal or is due to statistical fluctuations in the data. 

Several tests of statistical independence exist for both discrete and continuous distributions, including mutual information~\cite{Bishop:998831}, Hoeffding's D independence test~\cite{10.1214/aoms/1177730150} and distance correlation~\cite{10.1214/009053607000000505}. For simplicity, in the present work we focus on the use of the estimated mutual information (MI) $I$ of the measured probability distribution. MI encodes the exact property we want to test as it measures the difference between the joint distribution and the marginals:
\begin{align}
I(s,y)&=D_{\text{KL}}(p(s,y)||p(s)p(y))\\
&=\int \,ds\,dy\ p(s,y)\log \frac{p(s,y)}{p(s)p(y)}    \,,
\end{align}
where $D_{\text{KL}}(p,q)$ is the Kullback-Leibler divergence between two probability distributions, capturing how much information is lost when approximating the distribution $p$ with the distribution $q$. The MI thus captures how well one can approximate the joint distribution by the product of its marginals and it is trivial to show that it vanishes for independent variables. Conditional Independence can then be expressed as a vanishing MI conditioned on a given process
\beq
I(s,y|z)=\int \,ds\,dy\ p(s,y|z)\log \frac{p(s,y|z)}{p(s|z)p(y|z)} = 0\,.
\eeq
On the other hand, for the full dataset the possible mixture between the two processes encoded in $\pi\in[0,1]$ results in 
\beq
I(s,y)\geq 0\,,
\eeq
with the equality achieved when there is only one process or the two processes have the same probability distributions.

A very nice feature of the MI is that it has well behaved asymptotic properties in the limit of small MI and large sample size~\cite{mutual_info_asympt}. Thus, we can estimate it from the measured sample of $N$ events and obtain the p-value of said estimator $\hat{I}(s,y)$ under the null hypothesis $I(s,y)=0$. Assuming a two dimensional binning of $(s,y)$ with $d_{s}$ and $d_{y}$ the number of chosen bins per variable, the estimator $\hat{I}$ is a random variable that under the null hypothesis $I=0$ follows a Gamma distribution with shape parameter $\frac{(d_{s}-1)(d_{y}-1)}{2}$ and scale parameter $N$.

To estimate $\hat{I}$ we need to estimate $\hat{p}(s,y)$, with $\hat{p}(s)$ and $\hat{p}(y)$ obtained by marginalizing. We estimate $\hat{p}(s,y)$ through two-dimensional histogram event counts with the aforementioned $d_{s}$ and $d_{y}$ chosen bins. Because we are dealing with continuous variables, the use of binning introduces additional hyperparameters. In this work we bin $s$ and $y$ in such a way that each bin has a relative statistical uncertainty equal or lower than $1\%$. Other criteria for statistical independence that deal explicitly with continuous variables such as Hoeffding's D independence test or distance correlation could be used to avoid the introduction of binning at the expense of increased computational cost. We choose MI as it is straight forward to implement with a general signal-blind binning criteria and it suffices to establish the relevance of the strategy detailed in this work.

We emphasize that the role of \textsc{CWoLa} is to provide a one-dimensional observable $s(\vec{x})$ which can then be combined with $y$ to test for statistical independence. Once $s(\vec{x})$ is obtained, the rest of the test relies only on data without the need to introduce additional cuts or labels. If testing for statistical dependence between $\vec{x}$ and $y$ directly was feasible, then one would not need to introduce any learnable function. However, this is often not the case. One in general needs a high-dimensional $\vec{x}$ to ensure discriminative power between possible signals and the background, which is in turn converted by our method into statistical power to exclude statistical independence. On the other hand working directly with a high-dimensional set of features renders any statistical test problematic either due to the test being designed for two variables, as is the case for Hoeffding's D independence test and distance correlation, or due to the necessary density estimation suffering from the course of dimensionality as is the case for the mutual information test presented in this work.

The method relies on the assumption of conditional independence between $\vec x$ and $y$. In a realistic application this is not ensured, specially when considering highly-discriminative variables between the background and potential signals. The presence of correlation between $\vec{x}$ and $y$ will result in non-null MI for each process separately. Thus, the p-value obtained from the MI estimation will be merely testing for conditional independence, not the presence of a single process. In other words, the null hypothesis ceases to be equal to the background-only hypothesis. This challenge is already present in current implementations of \textsc{CWoLa}, with correlations resulting in loss of classification power. 

One possible strategy introduced in Ref.~\cite{Benkendorfer:2020gek} is to ensure that $s(\vec{x})$ is agnostic to the correlation between $\vec{x}$ and $y$ through the addition of a simulated background dataset during the training stage. In this approach, named Simulation Assisted Classification Without Labels (\textsc{SA-CWoLa}), the loss function is modified with an additional term that incorporates the simulation dataset. Following Ref.~\cite{Benkendorfer:2020gek}, we define the loss function as 
\begin{eqnarray}
    \mathcal{L}_{\text{SA-CWoLa}}[s] &=& -\left(\sum_{\vec{x}_{n}\in M^{\text{data}}_{1}} \log s(\vec{x}_{n}) + \sum_{\vec{x}_{n}\in M^{\text{data}}_{2}} \log \left(1-s(\vec{x}_{n})\right)\right)  \nonumber\\&-&\lambda \left(\sum_{\vec{x}_{n}\in M^{\text{sim.}}_{1}} \log \left(1-s(\vec{x}_{n})\right) + \sum_{\vec{x}_{n}\in M^{\text{sim.}}_{2}} \log s(\vec{x}_{n})\right)\,, \label{eq:loss}
\end{eqnarray}
that inverts the labelling in the simulation so as to penalize learning background differences between $M_{1}$ and $M_{2}$, with $\lambda$ the hyper-parameter that controls the relative importance of said penalization
\footnote{We always reweigh the events from both data and simulation during training in such a way that the each of the four subset of events $\{M^{\text{data}}_{1},M^{\text{data}}_{2},M^{\text{sim.}}_{1},M^{\text{sim.}}_{2}\}$ has the same total weight.}. Note that the specific choice of the loss function is not relevant as long as it ensures proper decorrelation. Similarly, the simulated dataset does not need to be perfect, it only needs to encode accurately enough the correlation between $\vec{x}$ and $y$ for the background process. Learning to ignore the correlations in the background guarantees that excluding the null hypothesis $I(s,y)=0$ corresponds to excluding the background-only hypothesis $p(s,y)=p(s,y|\pi=0)$ and not merely excluding conditional independence $p(s,y|B)=p(s|B)p(y|B)$. The main drawback of introducing decorrelation is that the learned function $s(\vec{x})$ ceases to be optimal and {looses} classification power. Thus, one should tune $\lambda$ with a given criteria that balances learning to decorrelate between $\vec{x}$ and $y$ for $B$ and learning to distinguish between $B$ and $S$ through discriminating between $M_{1}$ and $M_{2}$. 

In this proof-of-principle, we are satisfied with presenting results for fixed $\lambda$ that is large enough to ensure decorrelation in the simulation sample and at the same time small enough so that $s(\vec{x})$ is sensitive to the presence of signal in the measured sample and improves over the naive significance estimation $S/\sqrt{B}$. To verify that decorrelation is enforced, we follow Ref.~\cite{Benkendorfer:2020gek} and compute the Area-Under-Curve (AUC) for the $s(\vec{x})$ classifier for the $M_{1}$ and $M_{2}$ samples in the simulation dataset. If the AUC is approximately $0.5$, we have a classifier that is not better than a random classifier and thus it has learned to ignore any possible correlations between $\vec{x}$ and $y$. We emphasize that the sole purpose of the simulation dataset is to ensure decorrelation, and we never compare data to simulation to obtain a significance after training. This makes the test more robust against background mismodelling than other unsupervised methods for anomaly detection which avoid the use of anomaly cuts at the expense of performing data-to-simulation hypothesis tests such as Refs.~\cite{DAgnolo:2018cun, DAgnolo:2019vbw, dAgnolo:2021aun,Letizia:2022xbe, Krzyzanska:2022mto, Chakravarti:2021svb}. This is only possible because we assume that the simulations are precise enough to capture qualitative correlations between features in data. The simulator precision needed for decorrelation to be effective is considerably less than what is needed for a full multivariate comparison with measurements which often suffers both from systematic biases in the simulations and from the computational cost of achieving a given statistical precision.

\section{Application: LHC Olympics}\label{sec:application}

In order to demonstrate its power, we apply our method to the LHC Olympics R\&D labelled dataset~\cite{Kasieczka:2021xcg}. The dataset is comprised of dijet events from two different sources: SM quantum chromodynamics (QCD) processes (background $B$), and the production of a hypothetical new resonance $W'$ with mass $m_{W'}=3.5$ TeV, decaying to two intermediate particles $X$ and $Y$ with masses $m_{X}=500$ GeV and $m_{Y}=100$ GeV, which in turn both decay promptly to pairs of quarks producing two large-radius jets with a two-prong substructure (signal $S$). Our variable of interest $y$ is the reconstructed dijet invariant mass $m_{jj}$ of the two hardest (in $p_{T}$) jets in the event. Mimicking the selection criteria of Ref.~\cite{Benkendorfer:2020gek}, the selected events have a reconstructed dijet mass $m_{jj}\in[3.1,3.9]\TeV$. To perform \textsc{CWoLa}~\cite{Metodiev:2017vrx}, we define two orthogonal regions $M_{1}\equiv\{m_{jj}\in[3.3,3.7]\TeV\}$ and $M_{2}\equiv\{m_{jj}\in[3.1,3.3]\TeV\cup[3.7,3.9]\TeV\}$. In the following, we refer to $S$ and $B$ as the total number of Signal and Background events in $M_{1}\cup M_{2}$.

For anomaly score input features $\vec x$, we choose a set of variables based on the invariant masses and the first N-subjettiness ratios~\cite{Thaler:2010tr,Thaler:2011gf} of the two selected jets. Ordering the jets by mass, with $j=1$ being the heavier jet, our variables are
\beq 
\vec{x}=\{m_{1}-m_{2},m_{2},\tau_{21,1},\tau_{21,2}\} \,.
\eeq 

The correlation between $\vec{x}$ and $m_{jj}$ is mostly concentrated in the correlations between $\{m_{1},m_{2}\}$ and $m_{jj}$. To illustrate how important they are, we define
\begin{equation}
    \Delta^{ab}_{c}(m^{\text{bin}}_{jj}) = \frac{\mathbb{E}[m_{c}|m_{jj}\in m^{\text{bin}}_{jj},a]-\mathbb{E}[m_{c}|b]}{\mathbb{E}[m_{c}|b]}\,,
    \label{eq:Delta}
\end{equation}
where $a,b\in\{B,S\}$ and $c\in\{1,2\}$. $\Delta^{ab}_{c}(m^{\text{bin}}_{jj})$ represents the relative difference between the average of $m_{c}$ in a given $m_{jj}$ bin $m^{\text{bin}}_{jj}$ for process $a$ and the average of the same observable over the whole $m_{jj}$ range for process $b$. If $a=b$, this observable conveys the presence of a correlation between $m_{c}$ and $m_{jj}$ for a given process. For $a\neq b$, this observable conveys the difference in the $m_{c}$ probability distributions for the two processes and its dependence on $m_{jj}$. By comparing $\Delta^{BB}_{c}$ with $(S/B)\Delta^{SB}_{c}$ we can check whether the correlations between $m_{c}$ and $m_{jj}$ can obscure the differences between signal and background that \textsc{CWoLa} aims to learn. The prefactor $S/B$ is introduced to account for the fact that the data contains less signal than background and originates from comparing the full data distribution to the background-only distribution and separating the signal and the background contributions:
\begin{equation*}
    \Delta^{S+B,B}_{c}=\frac{B}{S+B}\Delta^{BB}_{c}+\frac{S}{S+B}\Delta^{SB}_{c}\approx \Delta^{BB}_{c}+(S/B)\Delta^{SB}_{c}\,.
\end{equation*}

\begin{figure}[h!]
 \vspace{-0.3cm}
\begin{center}
 \begin{tabular}{cc}
 \hspace*{-5mm}
 \includegraphics[width=0.5\linewidth]{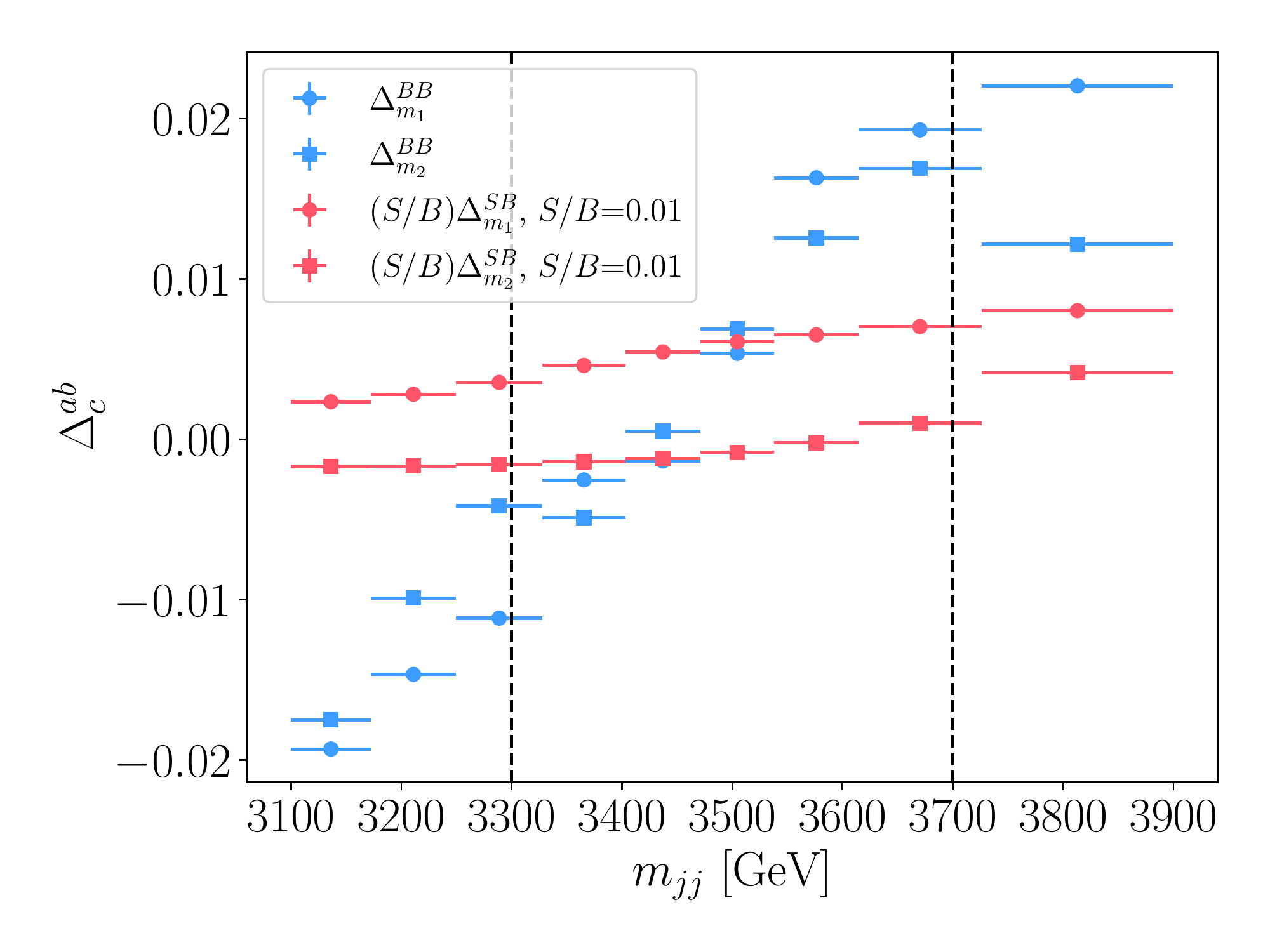}
 \hspace*{-0mm}
    \end{tabular}
\caption{Distribution of $\Delta^{ab}_{c}$ defined in Eq.~\eqref{eq:Delta} comparing correlations between $\{m_{1},m_{2}\}$ and $m_{jj}$ to the differences between $S$ and $B$ for the largest $S/B$ considered. Vertical error bars signify statistical uncertainties, while horizontal bars indicate $m_{jj}$ bin width. Vertical dashed lines denote boundaries between signal ($M_1$) and side-band ($M_2$) regions. See text for details. }\label{fig:mass_comparison}
\end{center}
\end{figure}

We show in Fig.~\ref{fig:mass_comparison} the resulting distributions for $m_{1}$ and $m_{2}$, with the largest $S/B$ considered in this work, $S/B=0.01$. We observe how the correlation between features for the background process, as evidenced by the monotonic increase of $\Delta^{BB}_{c}$ from negative to positive values towards larger $m_{jj}$, is sizable and crucially more pronounced compared to the $S/B$ weighted difference between $S$ and $B$ as traced by $\Delta^{SB}_{c}$. In other words, the correlation between $m_{c}$ and $m_{jj}$ in the background can easily mask the presence of a small signal. For lower $S/B$, the correlations become even more dominant and can lead to a strongly biased anomaly score. In addition, in our approach the correlations will also induce statistical dependence between $s(\vec x)$ and $M_{1,2}$ even in absence of a signal and thus jeopardize the validity of the null-hypothesis test.

In order to address this crucial issue, we follow Ref.~\cite{Benkendorfer:2020gek} and incorporate a set of simulated events into the training stage. As simulation, we consider the background provided in the labelled version of the Black Box 1 (BB1) dataset. We use BB1 as simulation to take advantage of the larger signal sample provided in the R\&D dataset. Using the loss function defined in Eq.~\eqref{eq:loss}, the classifier is then trained to distinguish $M_{1}$ and $M_{2}$ in data but not in simulation, obtaining a $s(\vec{x})$ that is agnostic to correlations between $\vec{x}$ and $y$ for QCD. As a classifier, we use a very similar set-up as in Ref.~\cite{Benkendorfer:2020gek}: we train a Neural Network composed of three hidden layers with 64 nodes each and \texttt{ReLU} activation function with a sigmoid function applied to the output to ensure $s(\vec{x})\in [0,1]$ for 20 epochs using the \texttt{ADAM}~\cite{adam} optimizer. The Neural Network is implemented in \texttt{PyTorch}~\cite{NEURIPS2019_9015}. We have trained our classifier using $k$-fold cross-validation with $k=10$ to avoid overfitting by ensuring that every $s(\vec{x}_{n})$ is obtained by combining the data point $\vec{x}_{n}$ with a classifier which has not seen $\vec{x}_{n}$ during training. We also perform several random weight initializations to ensure better convergence. At the end of training, we evaluate the AUC score between the $M_{1}$ and $M_{2}$ simulated samples to ensure that decorrelation is achieved.

\begin{figure}[h!]
 \vspace{-0.3cm}
\begin{center}
 \begin{tabular}{cc}
 \hspace*{-5mm}
 \includegraphics[width=.85\linewidth]{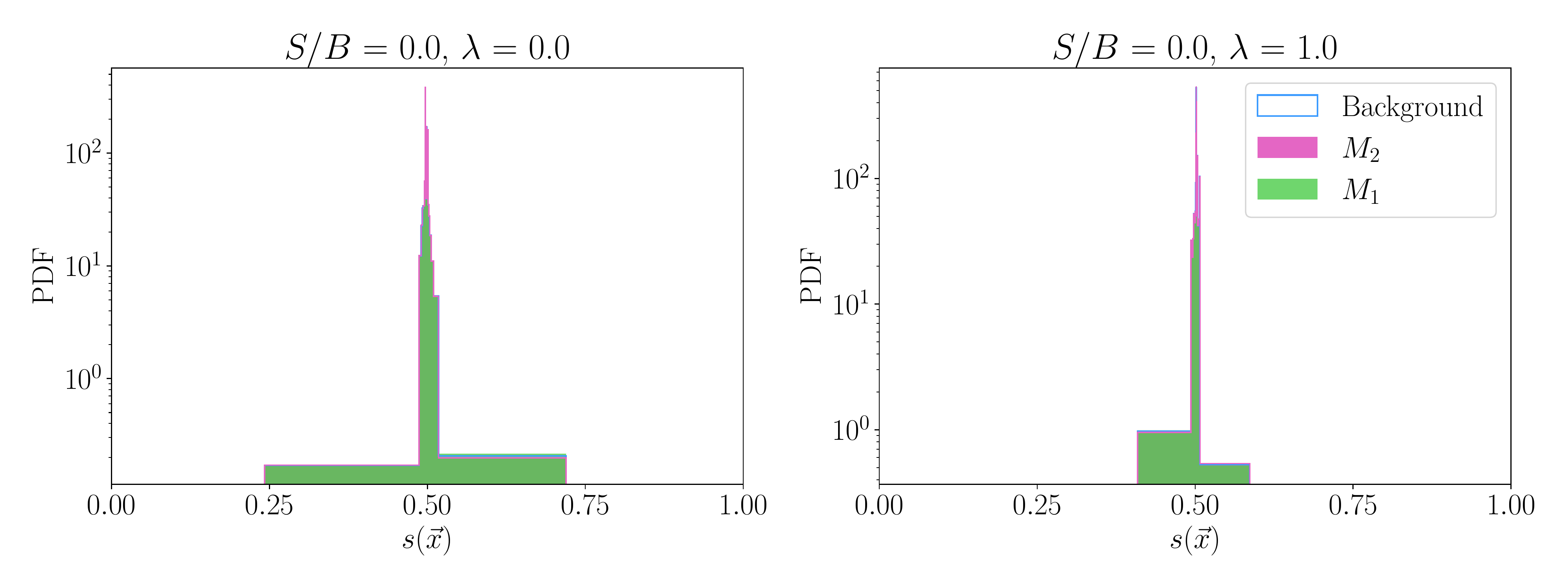}
 \hspace*{-0mm}
 \\
  \hspace*{-5mm}
 \includegraphics[width=.85\linewidth]{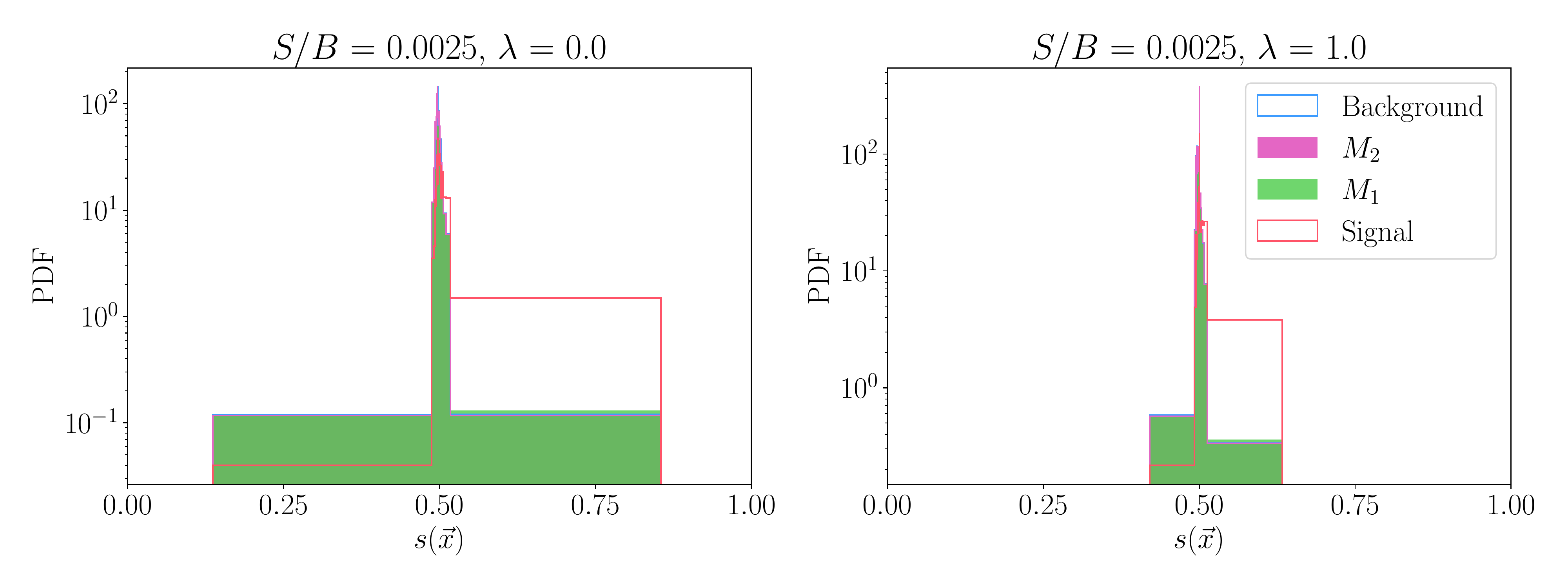}
 \hspace*{-0mm}
 \\
  \hspace*{-5mm}
 \includegraphics[width=.85\linewidth]{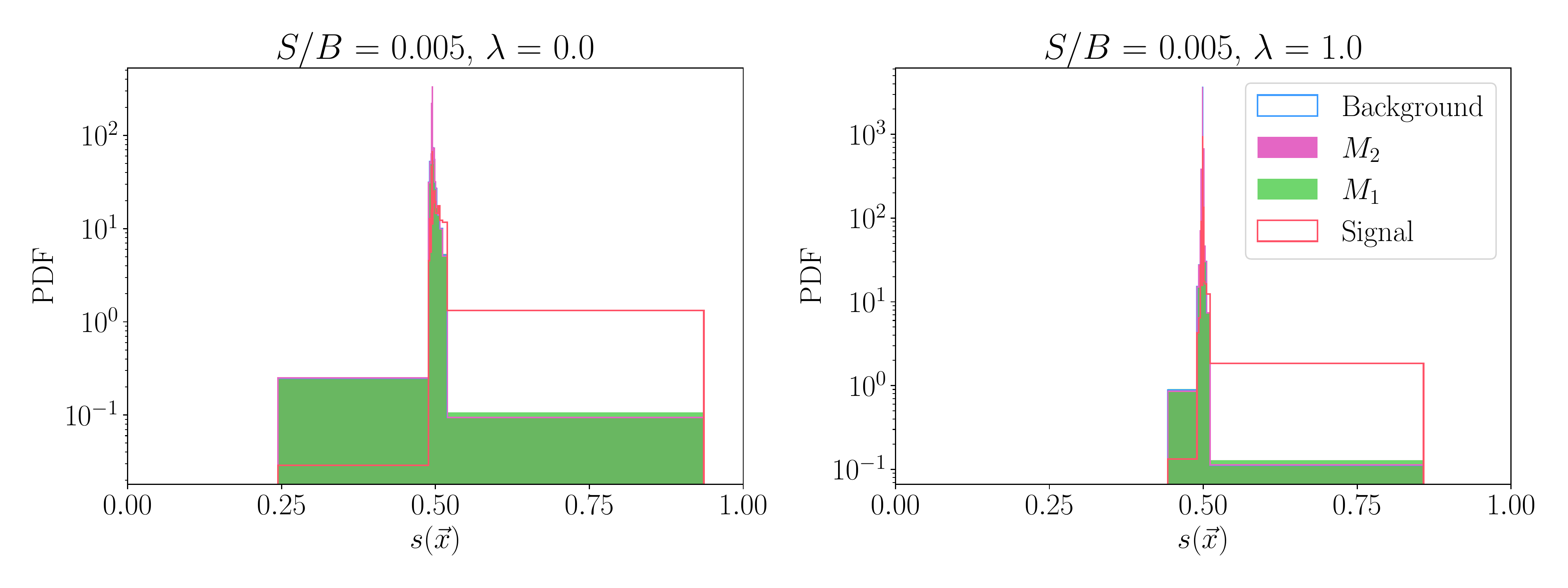}
 \hspace*{-0mm}
 \\
  \hspace*{-5mm}
 \includegraphics[width=.85\linewidth]{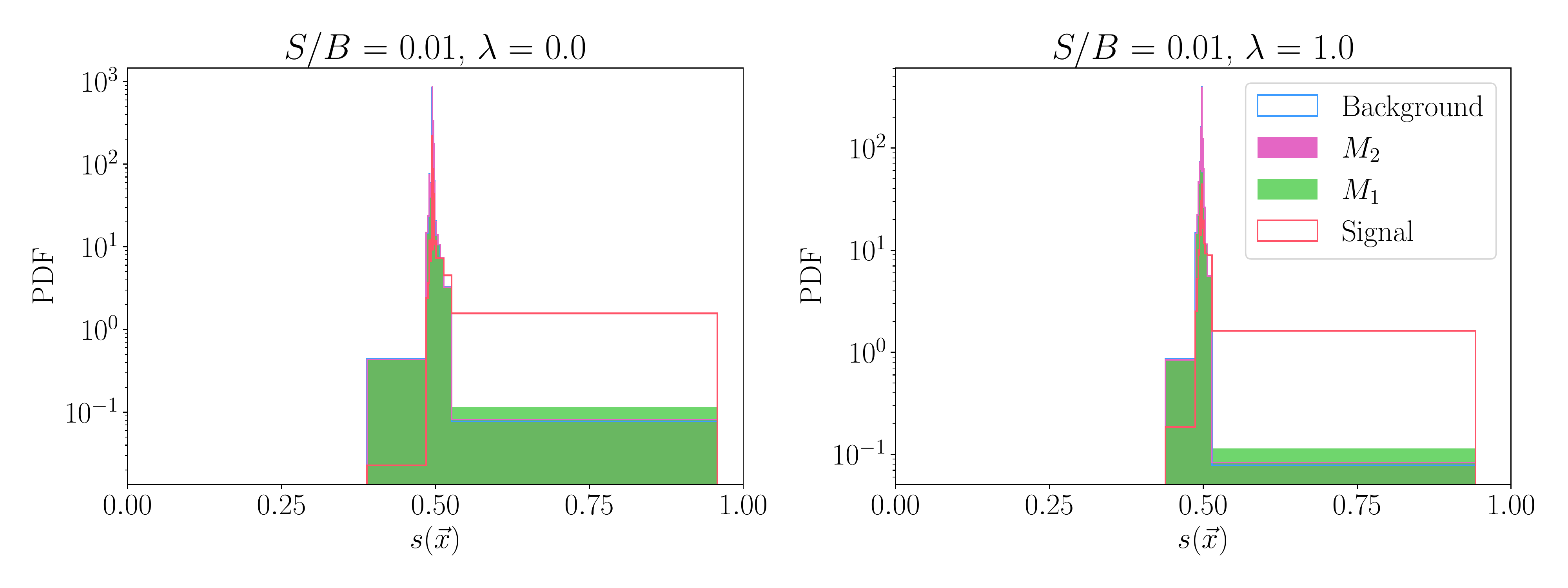}
 \hspace*{-0mm}
    \end{tabular}
\caption{Anomaly score $s(\vec{x})$ probability density function (PDF) after training for different event labellings. Each row corresponds to a given $S/B$ and each column to a given $\lambda$. Non-uniform binning ensures that when considering the full dataset each bin has a relative statistical uncertainty equal or lower than $1\%$, resulting in between 15 and 25 bins per plot. The data is shown both labelled according to the (observable) values of $m_{jj}$ (defining $M_{1}$ and $M_{2}$) as well as according to the (unobservable) truth labels (background and signal).}\label{fig:s_mjj_distributions_s_over_b}
\end{center}
\end{figure}

We show in Fig.~\ref{fig:s_mjj_distributions_s_over_b} the learned $s(\vec{x})$ for different $S/B$ with $B=250$k and $\lambda = \{0.0, 1.0\}$. In each plot, the binning is chosen in such a way that each bin has a relative statistical uncertainty lower than or equal to $1\%$. This choice ensures good performance of the density estimation needed for the hypothesis test. We can appreciate how as $S/B$ increases, $s(\vec{x})$ goes from being mostly centered around $s=0.5$ to yielding higher $s$ values, indicating improved learning of signal features. However, the binning choice obscures somewhat how much the background and signal are separated (within the highest $s$ bin), as the whole signal is grouped together with the necessary background events to obtain a $1\%$ statistical uncertainty. 
In absence of correlation mitigation (for $\lambda=0$), anomaly score bias causes clearly unbalanced classification of events in $M_1$ and $M_2$ even in absence of any signal. The introduction of $\lambda = 1$ causes the events to be even more centered around $0.5$, specially for low $S/B$. However, more importantly, $\lambda \ge 0$ forces the training to ignore possible correlations between $\vec{x}$ and $y$ for (simulated) background and thus approaches a random classifier for $S/B \to 0$.  

\begin{figure}[h!]
 \vspace{-0.3cm}
\begin{center}
 \begin{tabular}{cc}
 \hspace*{-5mm}
 \includegraphics[width=0.5\linewidth]{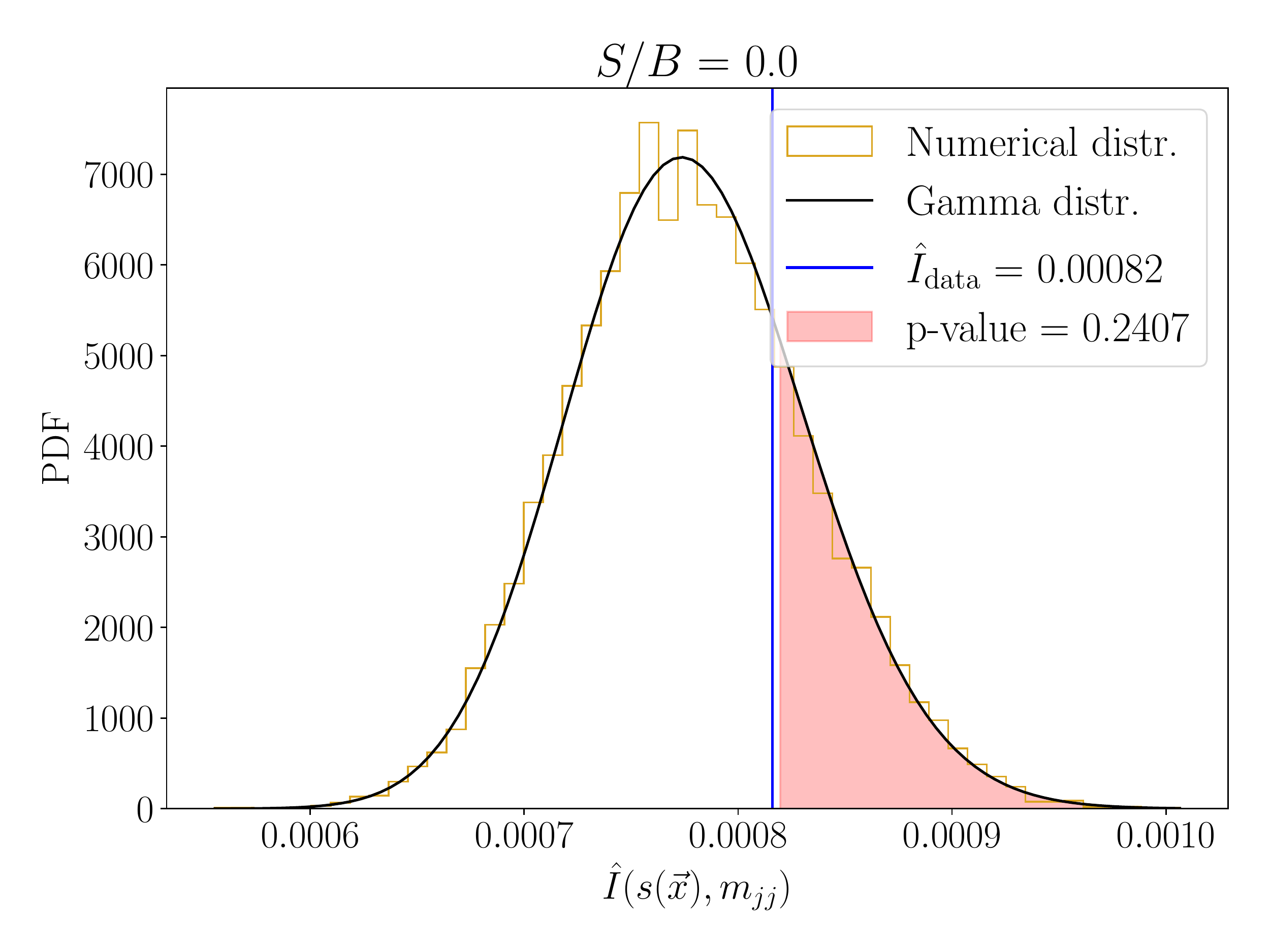}
 \hspace*{0mm}
 \includegraphics[width=0.5\linewidth]{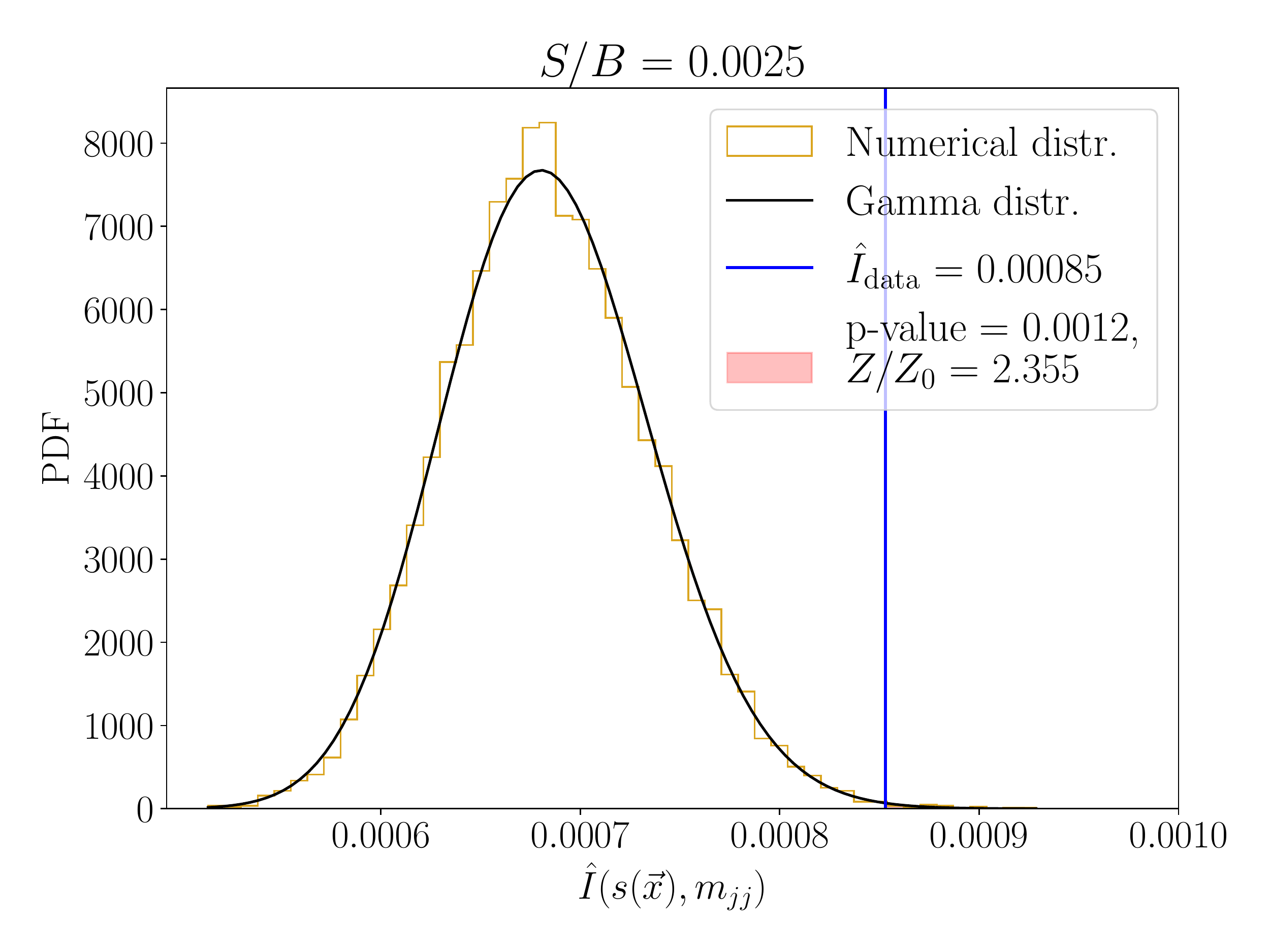}
  \hspace*{0mm}
 \\
 \includegraphics[width=0.5\linewidth]{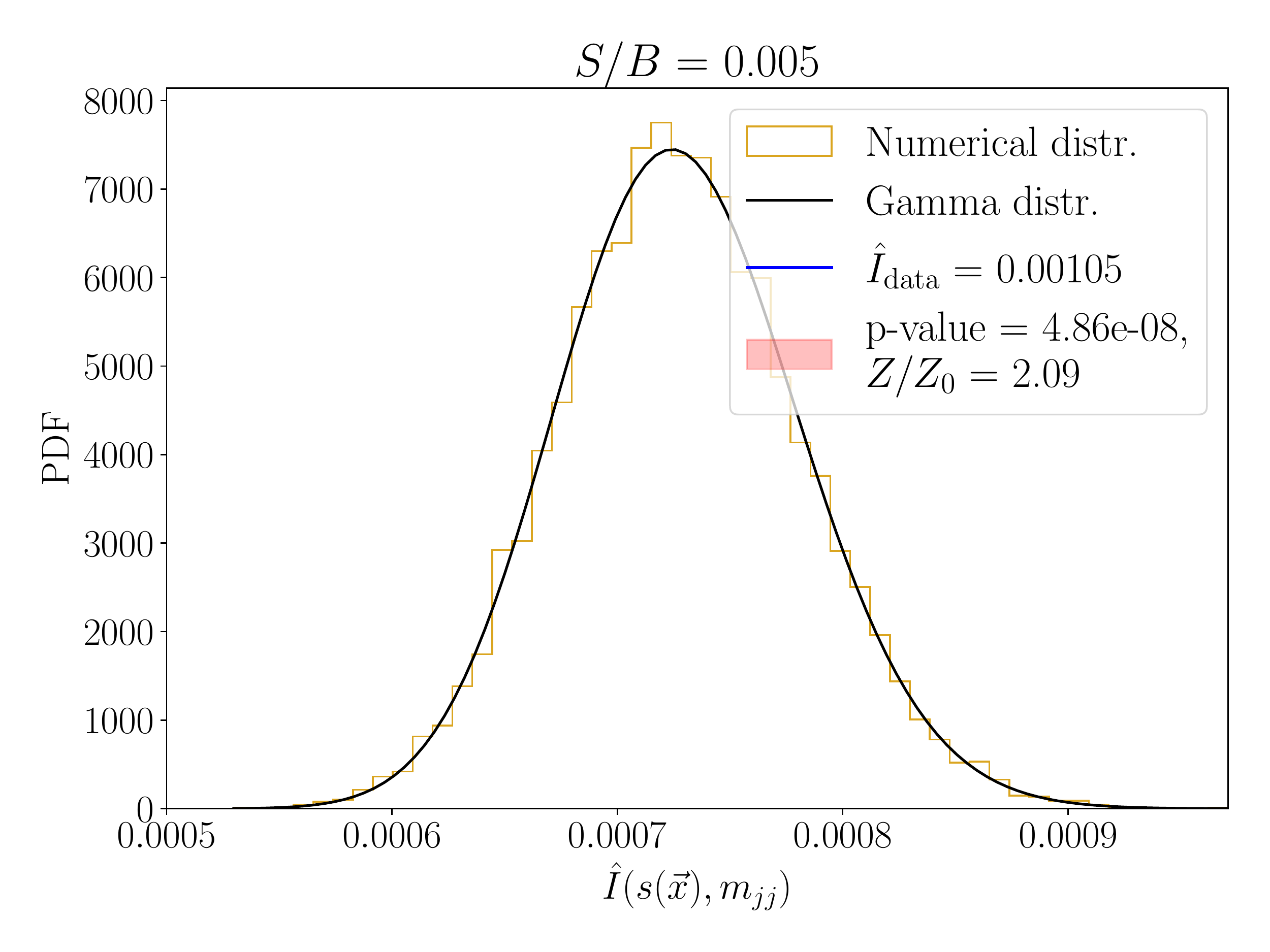}
 \hspace*{0mm}
 \includegraphics[width=0.5\linewidth]{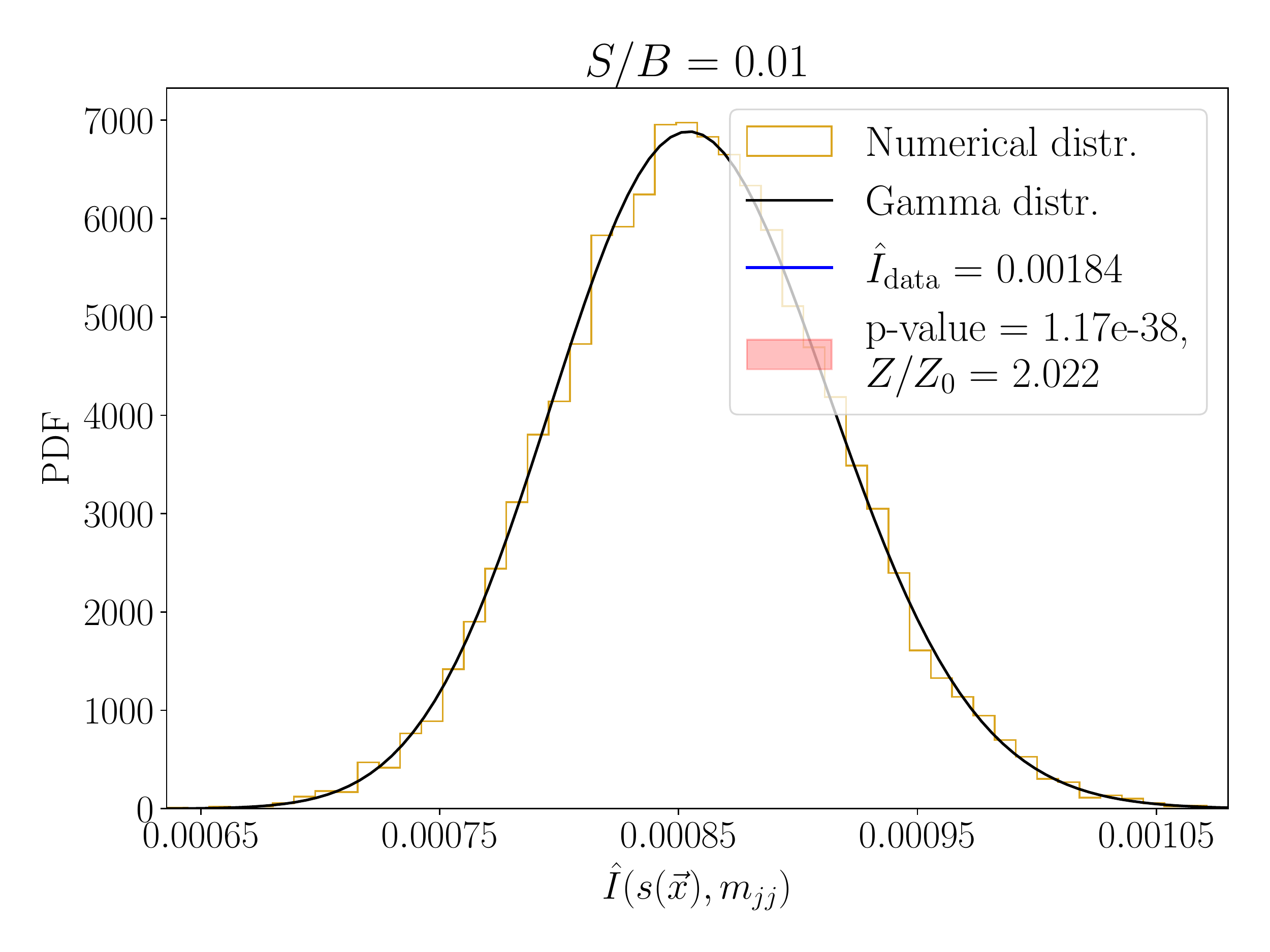}
    \end{tabular}
\caption{PDFs of estimated mutual information, its numerical distribution under the null hypothesis estimated through resampling, and its asymptotic distribution under the null hypothesis, for different considered benchmark datasets. Each plot corresponds to a different choice of $S/B$ with $\lambda = 1$. We combine the estimated $\hat{I}$ with their asymptotic distribution to obtain the resulting p-values. When $S/B > 0$, we also compute the discovery significance $Z=\Phi^{-1}(1-p)$, where $\Phi$ is the unit Gaussian cumulative distribution function, and compare it to $Z_{0}=S/\sqrt{B}$.}\label{fig:I_distributions}
\end{center}
\end{figure}

As expected, the impact of $\lambda$ is even more pronounced when testing for statistical independence. For each dataset of $\{s,m_{jj}\}$ values, we estimate mutual information and obtain the p-value associated with the null hypothesis as detailed in Section~\ref{sec:method}. We show in Fig.~\ref{fig:I_distributions} the resulting estimated $\hat{I}_{\mathrm{data}}$ and their corresponding p-values. We also ran a series of pseudo-experiments to verify that the asymptotic limit is appropriate. We only present results for $\lambda = 1$ because for $\lambda = 0$ we are able exclude $I(s,y)=0$ for all $S/B$ with very high confidence (p-value$\,< 10^{-14}$). This, as detailed in Section~\ref{sec:method}, is because we are excluding conditional independence in the background process due to correlations between $\vec{x}$ and $y$. This is specially important for $S/B=0$, where the effect of correlations can mislead \textsc{CWoLa} to falsely exclude the background-only hypothesis.

We observe that for $\lambda = 1$ the proposed test has the required behavior: for $S/B=0$ the test yields results consistent with statistical independence, while an increase of $S/B$ leads to an increasingly strong exclusion of the null hypothesis. The use of \textsc{SA-CWoLa} thus ensures that we can identify the null hypothesis with the background-only hypothesis. For $S/B > 0$, we also compute the discovery significance $Z=\Phi^{-1}(1-p)$, where $\Phi$ is the unit Gaussian cumulative distribution function, and compare it to the naive counting significance $Z_{0}=S/\sqrt{B}$. We observe how our method presents an increased discovery significance even compared to the case of perfect (up to statistical fluctuation) knowledge of background yields.

Overall, Fig.~\ref{fig:I_distributions} shows how $\hat{I}$ can be used to infer whether the data presents a deviation from the null hypothesis, defined as the case where a single process (or an $m_{jj}$ independent mixture of processes) is present for which $s$ and $m_{jj}$ are independent. This is the analogous to the p-value obtained using the Bump Hunt in a ``traditional" implementation of \textsc{CWoLa}. However, contrary to existing approaches, here there is no selection cut to be optimized, and no extrapolation of the background into the signal region is required. 

In principle the method could learn the likelihood-ratio between signal and background, which is the optimal test-statistic. However, the presence of the decorrelation term in Eq.~\eqref{eq:loss} reduces the optimality of $s$ and consequently $\hat{I}$. This can be seen by the decrease in $Z/Z_{0}$ as $S/B$ increases. When $S/B$ is large enough, $s(\vec{x})$ will ignore the small correlations in each individual process even for $\lambda = 0.0$. A non-null $\lambda$ will thus only worsen the performance of the algorithm. However, as we are interested in low $S/B$ cases where anomaly detection is useful, a more conservative approach which is robust to correlations even when there is no signal present is warranted.

In the previous paragraphs, we have shown how the proposed method yields an appropriate test statistic which is different from existing approaches.  In Table~\ref{tab:comparison} we provide a significance comparison of the proposed method to two traditional implementation of \textsc{CWoLa} which we denote as ``Anomaly cuts" and ``Bump Hunt". The former, implemented e.g. in Ref.~\cite{Finke:2022lsu}, assumes that no signal is present in the side-band region $M_{2}$ and estimates the total number of background events in the signal region $M_{1}$ through the use of cuts on the anomaly score. For a fixed efficiency in the side-band region $\epsilon_{2}$, the estimated background event yield in signal region is $\epsilon_{2}N_{1}$. If the measured efficiency in the signal region $\epsilon_{1}$ is larger than $\epsilon_{2}$, then there is an excess of events with a significance of 
\begin{equation*}
    Z=\begin{cases}
    \frac{(\epsilon_{1}-\epsilon_{2})N_{1}}{\sqrt{\epsilon_{2}(N_{1}+N_{2})}},& \quad \text{if } \epsilon_{1} \geq \epsilon_{2} \,,\\[2ex]
    0,& \quad \text{if } \epsilon_{1} < \epsilon_{2}\,.
      \end{cases}
\end{equation*}

Similarly, the Bump Hunt method also assumes no signal is present in $M_{2}$ but estimates the background event yield in $M_{1}$ differently.  In this approach, the background $m_{jj}$ distribution is explicitly modelled and a Profile Likelihood Ratio fit of the number of signal events in $M_{1}$ is performed. We follow Ref.~\cite{Benkendorfer:2020gek} and model the background distribution for $m_{jj} \in [3.1,3.9]$ TeV as
\begin{equation*}
    \frac{d\sigma}{dm_{jj}} = \frac{p_{0}\left(1-\frac{m_{jj}}{\sqrt{s}}\right)^{p_{1}}}{\left(\frac{m_{jj}}{\sqrt{s}}\right)^{p_{2}+p_{3}\log \frac{m_{jj}}{\sqrt{s}}}}\,,
\end{equation*}
where $\sqrt{s}$ is the center-of-mass energy and $p_{i}$ are parameters to be fitted from the $m_{jj}$ distribution with the signal region masked. Once the fit is performed, the expected number of background events in $M_{1}$ $b$ is estimated from the integral of the background distribution over the signal region.We define the Likelihood function in the signal region as
\begin{equation*}
    \mathcal{L}(s,\theta)=\mathcal{P}(N_{1}|s+b+\theta)\mathcal{N}(\theta|0,\sigma)\,,
\end{equation*}
where $\mathcal{P}$ is the Poisson probability mass function of measuring $N_{1}$ events, $\mathcal{N}$ is the Normal probability density, $\theta$ is a nuisance parameter for background mismodelling and $\sigma$ is the background yield error propagated from the $p_{i}$ fit. From this Likelihood we build the usual test statistic
\begin{equation*}
    q_{0}=\begin{cases}
    - 2 \text{ Ln } \mathcal{L}(0,\hat{\hat{\theta}})/\mathcal{L}(\hat{s},\hat{\theta}),& \quad \text{if } \hat{s} \geq 0\,,\\[2ex]
    0,& \quad \text{if } \hat{s} < 0\,,
      \end{cases}
\end{equation*}
where $\hat{s},\hat{\theta}$ are the maximum likelihood estimates of $s$ and $\theta$ and $\hat{\hat{\theta}}$ is the maximum likelihood estimate of $\theta$ when keeping $s$ fixed to 0. The resulting significance is $Z=\sqrt{q_{0}}$. We implement the Bump Hunt by itself and in conjunction with the use of cuts in the anomaly score to enhance the efficiency as would be done in a resonance search.

From Table~\ref{tab:comparison} we observe how in every case the introduction of $\lambda > 0$ reduces the significance. However, it does not imply resilience against spurious signals for all strategies. Both traditional methods are highly dependent  on the arbitrary $\epsilon_{2}$ choice, showing the appearance of spurious significance at $S/B = 0$ for certain values. In general, we observe that our method is better suited for smaller $S/B$ than existing methods. This is mainly because it does not discard potential signal events. The Bump Hunt with no cuts also does this, but its significance is lower for non-null $S/B$ (for the extreme $S/B = 0.01$, its significance is even lower than $S/\sqrt{B}$ due to the presence of the nuisance parameter). From this comparison, we assess that without clear criteria for an optimal anomaly cut ($\epsilon_2$), the mutual information-based method performs better for low to null $S/B$ whereas the cut and count based methods outperform the mutual information test for larger $S/B$.

Another benefit of our model compared to traditional searches is scalability with sample size. If decorrelation is ensured, the larger the sample size the more powerful the method. This is certainly not true for the Bump Hunt, where the background modelling is an inherent approximation which necessarily introduces bias for large enough sample sizes. 
Conversely, for smaller datasets our method takes advantage of the full dataset in a better way than through the use of fixed anomaly cuts. The asymptotic approximation for the mutual information CDF, which is vital for the test, has been shown numerically~\cite{mutual_info_asympt} to be valid for $N > 50$ events and true mutual information $I \leq 0.14$, conditions that we expect to always be satisfied in a realistic anomaly search at the LHC.

\begin{table}[t]
\renewcommand{\arraystretch}{1}
\centering
\begin{tabular}{|c|c|c|c|c|}
\hline\hline
 Significance & $S/B$ = 0.0 & $S/B$ =  0.0025 & $S/B$ =  0.005 & $S/B$ =  0.01 \\ \hline
$S/\sqrt{B}$ & 0.0 & 1.29 & 2.55 & 6.40 \\ \hline
Mutual Info $\lambda = 0.0$ & 6.40 &  7.04 &  7.58 & 14.1 \\
Mutual Info $\lambda = 1.0$ & 0.70 & 3.03 &  5.33 & 13.0 \\ \hline 
Anomaly cuts $\epsilon_{2} = 0.1$, $\lambda = 0.0$ & 3.35 &  4.78 &  6.27 & 11.6   \\
Anomaly cuts $\epsilon_{2} = 0.1$, $\lambda = 1.0$ & 2.48 &  2.26 &  4.49 & 10.0   \\\hline
Anomaly cuts $\epsilon_{2} = 0.01$, $\lambda = 0.0$ & 2.26 &  4.62 & 10.1 & 27.0   \\
Anomaly cuts $\epsilon_{2} = 0.01$, $\lambda = 1.0$ & 0.55 &  1.66 & 10.7 & 27.1   \\\hline
Anomaly cuts $\epsilon_{2} = 0.001$, $\lambda = 0.0$ & 1.39 & 10.3 & 17.9 & 34.2  \\
Anomaly cuts $\epsilon_{2} = 0.001$, $\lambda = 1.0$ & 0. &  0.57 & 13.6 & 37.0   \\\hline
Bump Hunt & 0.95 &  1.97 &  2.74 &  5.30  \\\hline
Bump Hunt $\epsilon_{2} = 0.1$, $\lambda = 0.0$ & 6.41 &  9.26 & 10.92 & 19.5   \\
Bump Hunt $\epsilon_{2} = 0.1$, $\lambda = 1.0$ & 3.81 &  4.35 &  6.93 & 16.0   \\\hline
Bump Hunt $\epsilon_{2} = 0.01$, $\lambda = 0.0$ & 4.77 &  6.96 & 14.2 & 34.7   \\
Bump Hunt $\epsilon_{2} = 0.01$, $\lambda = 1.0$ & 0.97 & 2.53 & 14.0 & 35.0   \\\hline
Bump Hunt $\epsilon_{2} = 0.001$, $\lambda = 0.0$ & 2.98 & 12.3 & 20.0 & 35.8  \\
Bump Hunt $\epsilon_{2} = 0.001$, $\lambda = 1.0$ & 0.29 &  1.60 & 15.9 & 38.7   \\
\hline \hline
\end{tabular}
\caption{Significances obtained with different strategies for different $S/B$ ratios, see text for details.}
\label{tab:comparison}
\end{table}

\section{Discussions and outlook}\label{sec:discussion}

In this work, we have presented a novel strategy to quantify the sensitivity of a specific anomaly detection technique, Simulation-Assisted Classification Without Labels, by testing for statistical independence of the learned $\{s(\vec{x}),y\}$ samples. We have shown that as long as one can rely on \textsc{SA-CWoLa} to enforce conditional independence of the background processes $p(s,y|B)=p(s|B)p(y|B)$, the null hypothesis of statistical independence is equivalent to the background-only hypothesis. Thus, testing for statistical independence in the observed data corresponds to testing for the background-only hypothesis.

As a proof of principle, we have considered mutual information as a test statistic. MI has a known asymptotic distribution under the null hypothesis for binned data and has low computational cost. We have tested our method with LHC Olympics datasets and have shown that the test statistic yields the expected behavior. Most importantly our proposed test statistic provides a clear statement on the presence of signal, i.e. is capable of correctly yielding a no-signal response. This opens the door to testing for new physics in LHC datasets without the need for anomaly score cuts, as well as reducing the need for accurate background modelling. 

Possible extensions of the present work could consider other tests for statistical independence such as Hoeffding's D independence test or distance correlation, which can be applied on unbinned $s(\vec{x})$ and $y$, at the expense of increased computational cost. Similarly, other methods for anomaly score training and decorrelation of features could be explored. Employing  the most suitable classification and decorrelation methods can be model and dataset dependent, and has not been the main focus of this work. 

Another possibility is to assume that no signal populates the side-bands and identify $p(\hat{s}|M_{2})=p(\hat{s}|z=0)$. This opens the door for an optimal analysis since one can now perform a template fit in the signal region~\cite{Nachman:2019dol}. However, it also potentially introduces additional uncertainties and/or biases due to background modelling that the present method avoids. We leave a more complete study in this direction for future work.

Regarding other physics applications, we emphasize that by dispensing with the need for explicit functional background modelling, our test is especially useful for anomaly detection applications that do not search for predetermined (modelled) signal shapes~\cite{Shanahan:2022fzy}, such as invariant mass resonances as in e.g. Refs.~\cite{Alvarez:2019knh,Finke:2022lsu}. However, it could also be applied in supervised searches as an additional cross-check to control bias due to modelling at the expense of loss of optimality.

\section*{$\footnotesize\text{\ding{117} \ding{117} \ding{117}}$\ \  Acknowledgments\ \  $\footnotesize\text{\ding{117} \ding{117} \ding{117}}$}
The authors acknowledge the financial support from the Slovenian Research Agency (grant No. J1-3013 and research core funding No. P1-0035). MS is grateful to the Mainz Institute for Theoretical Physics (MITP) of the Cluster of Excellence PRISMA${}^+$ (Project ID 39083149) and to GGI for their hospitality, support and useful discussions.

\bibliographystyle{apsrev4-1}
\bibliography{refs}

\end{document}